%% file: ms.tex
\documentclass[12pt]{article}
\usepackage{times}
\usepackage{geometry}
\usepackage[utf8]{inputenc}
\usepackage{enumitem,amssymb}
\usepackage{ragged2e}
\newlist{thematic}{itemize}{8}
\setlist[thematic]{label=$\square$}
\usepackage{pifont}
\newcommand{\cmark}{\ding{51}}%
\newcommand{\done}{\rlap{$\square$}{\raisebox{2pt}{\large\hspace{1pt}\cmark}}%
\hspace{-2.5pt}}

\usepackage{pslatex}
\usepackage[square,comma,numbers]{natbib}
\usepackage{color}
\usepackage{graphicx}
\usepackage{upgreek}
\usepackage{amsmath, amsthm, amssymb}

\usepackage{sectsty}
\sectionfont{\fontsize{12}{15}\selectfont}

\input macros.tex

\begin{document}
        
\raggedright

\huge
Astro2020 Science White Paper \linebreak

First Stars and Black Holes at Cosmic Dawn with Redshifted 21-cm Observations \linebreak
\normalsize

\noindent \textbf{Thematic Areas:} \hspace*{60pt} $\square$ Planetary Systems \hspace*{10pt} $\square$ Star and Planet Formation \hspace*{20pt}\linebreak
$\square$ Formation and Evolution of Compact Objects \hspace*{31pt} $\done$ Cosmology and Fundamental Physics \linebreak
  $\square$  Stars and Stellar Evolution \hspace*{1pt} $\square$ Resolved Stellar Populations and their Environments \hspace*{40pt} \linebreak
  $\done$    Galaxy Evolution   \hspace*{45pt} $\square$             Multi-Messenger Astronomy and Astrophysics \hspace*{65pt} \linebreak
  
\textbf{Principal Author:}

Name: Jordan Mirocha
 \linebreak						
Institution:  McGill University
 \linebreak
Email: jordan.mirocha@mcgill.ca
 \linebreak
%Phone: 
% \linebreak
 
\textbf{Co-authors:} (names and institutions)
  \linebreak
  Daniel Jacobs (Arizona State), Josh Dillon (UC-Berkeley), Steve Furlanetto (UCLA), Jonathan Pober (Brown), Adrian Liu (McGill), James Aguirre (UPenn), Yacine Ali-Ha\"imoud (NYU), Marcelo Alvarez (UC-Berkeley), Adam Beardsley (Arizona State), George Becker (UC-Riverside), Judd Bowman (Arizona State), Patrick Breysse (CITA), Volker Bromm (UT-Austin), Jack Burns (CU-Boulder), Xuelei Chen (Chinese Academy of Sciences), Tzu-Ching Chang (JPL), Hsin Chiang (McGill), Joanne Cohn (UC-Berkeley), David DeBoer (UC-Berkeley), Cora Dvorkin (Harvard), Anastasia Fialkov (Sussex), Nick Gnedin (Fermilab), Bryna Hazelton (Washington), Masui Kiyoshi (MIT), Saul Kohn (Vanguard Group), Leon Koopmans (Kapteyn Astronomical Institute), Ely Kovetz (Ben-Gurion), Paul La Plante (UPenn), Adam Lidz (UPenn), Yin-Zhe Ma (KwaZulu-Natal), Yi Mao (Tsinghua), Andrei Mesinger (Scuola Normale Superiore), Julian Mu\~noz (Harvard), Steven Murray (Arizona State), Aaron Parsons (UC-Berkeley), Jonathan Pritchard (Imperial College London), Jonathan Sievers (McGill), Eric Switzer (Goddard), Nithyanandan Thyagarajan (NRAO), Eli Visbal (Flatiron), Matias Zaldarriaga (Institute for Advanced Study)
  \linebreak
  
\textbf{Abstract  (optional):}
The ``cosmic dawn'' refers to the period of the Universe's history when stars and black holes first formed and began heating and ionizing hydrogen in the intergalactic medium (IGM). Though exceedingly difficult to detect directly, the first stars and black holes can be constrained indirectly through measurements of the cosmic 21-cm background, which traces the ionization state and temperature of intergalactic hydrogen gas. In this white paper, we focus on the science case for such observations, in particular those targeting redshifts z $\gtrsim$ 10 when the IGM is expected to be mostly neutral. 21-cm observations provide a unique window into this epoch and are thus critical to advancing first star and black hole science in the next decade.

\pagebreak

\RaggedRight

%%
% INTRO
%%
\section{Introduction} \vspace{-12pt}
The search for the first stars and black holes to form after the Big Bang is a key focus of the galaxy evolution community today. This has motivated new observations targeting 21-cm emission from neutral hydrogen atoms at high redshifts ($z \gtrsim 6$), which broadly traces the distribution of matter and energy at that time and thus holds significant discovery potential. 

Theoretical models of this epoch generally predict the formation of massive stars capable of regulating galaxy formation near and far through intense ultraviolet (UV) radiation and powerful supernovae explosions. Black hole (BH) remnants of the first stars may be the seeds of super-massive black holes (SMBHs) powering bright quasars as early as redshift $z \sim 7$. Little is known about these objects, unfortunately, leaving critical gaps in our understanding of galaxy formation. When and how do the first stars and BHs form, and how do their properties and abundance affect galaxy evolution?

Though ``stellar archaelogy'' and high-$z$ supernova searches are promising probes of the first stars, in this white paper, we focus solely on the first stars and BHs science enabled by low-frequency radio observations targeting 21-cm emission from before reionization (i.e., $z \gtrsim 10$). Such observations \textit{directly} probe the high-$z$ intergalactic medium (IGM), and thus \textit{indirectly} probe sources capable of heating and ionizing intergalactic gas. Given that the first star forming regions are too faint to be detected directly in the near term, \textit{21-cm observations are likely the only way to constrain the first generations of stars and BHs in the Universe in the next decade.}

The recent report of a 78 MHz feature in the sky-averaged signal by EDGES \citep{Bowman2018} has received much attention due to its anomalous depth, which may be evidence of non-gravitational interactions between baryons and dark matter \citep{Barkana2018,Munoz2018,Berlin2018,Kovetz2018} or excess power in the Rayleigh-Jeans tail of the cosmic microwave background \citep[CMB;][]{Feng2018,Pospelov2018,Sharma2018,Fraser2018}, and places interesting limits on other classes of dark matter and dark energy models \citep[e.g.,][]{Hill2018,Safarzadeh2018,Lidz2018,Schneider2018}. However, if correct, the EDGES signal also represents the earliest evidence of star formation in the Universe (redshift $z \gtrsim 18$), and constrains the efficiency of star and BH formation in small halos at $10 \lesssim z \lesssim 30$ \citep{EwallWice2018,Mirocha2019,Madau2018,Schauer2019,Kaurov2018}. The diversity of responses to the EDGES result illustrates the rich information content of the 21-cm background and thus serves as an important reminder: observations of hydrogen through the dark ages and the cosmic dawn probe the Universe at a time and density scale where novel physics has an outsize influence, and, as a result, yet more surprises may be in store for those able to make such challenging measurements.

The plan for this white paper is as follows. We outline the opportunities for 21-cm cosmology in the next decade and the broader scientific context in Sections \ref{sec:opportunities} and \ref{sec:context}, respectively. In Section \ref{sec:advances}, we discuss the advances needed to enable first star and BH science at redshifted 21-cm wavelengths in the next decade. For more information regarding 21-cm efforts targeting reionization itself, synergies between 21-cm observations and galaxy surveys during reionization, intensity mapping with other lines, and cosmology enabled by 21-cm observatories, see white papers by Furlanetto et al., Burns et al., Alvarez et al., and Liu et al.

%Danny: hydrogen through the dark ages and the cosmic dawn probes the early universe at a time and density scale where novel physics has an outsize influence.  Energetic input by the first stars, a solid prediction backed by much evidence, is arguably the most favored theory and remains the focus of much investigation. However with few competing observables at the Cosmic Dawn, the unexpected might very well be the order of the day and a worthy reward to a challenging observation.  

%%
% OPPORTUNITIES
%%
\vspace{-12pt}
\section{Opportunities for the Next Decade} \label{sec:opportunities} \vspace{-12pt}
Current constraints on the properties of the first stars and BHs are very limited. For example, CMB observations constrain the timing and duration of reionization through the measures of the Thomson scattering optical depth \citep[e.g.,][]{Robertson2015,Bouwens2015c} and the kinetic Sunyaev-Z'eldovich (kSZ) effect \citep[e.g.,][]{Miranda2017}, respectively, and thus constrain the ionizing emissions of high-$z$ sources in aggregate. Alternatively, limits on the amount of star formation and BH accretion at high-$z$ can be obtained by appealing to unresolved emission in the $z=0$ near-infrared background (NIRB) and X-ray background (XRB), respectively \citep[e.g.,][]{Fernandez2013,Salvaterra2012}. However, each approach yields constraints that are difficult to interpret: the CMB optical depth constrains the volume- and time-averaged ionizing emissivitiy of galaxies, while the NIRB and XRB suffer from mixed temporal and spectral information. As a result, contributions from ``known'' sources must be modeled exquisitely in order to enable constraints on the faintest, highest redshift sources.

Observations of the 21-cm line from hydrogen are advantageous in that the mapping between frequency and time (or redshift) is 1:1, modulo small ``redshift space distortions'' (RSD) due to peculiar velocities along the line of sight. As a result, one can in principle reconstruct the history of star and BH formation with measurements at a series of frequencies. The challenge is the need to observe at low radio frequencies, $\nu_{\mathrm{obs}} = 1.4 \ \mathrm{GHz} / (1 + z)$, where the galactic foreground emissions grow stronger as $\sim \nu^{-2.5}$. Though bright, the foreground is spectrally smooth whereas the 21-cm signal has spectral structure. As a result, separation of signal from foreground is in principle possible with precisely calibrated experiments and careful analyses \citep[e.g.,][]{Shaver1999,Pritchard2010a,Harker2012}. In the next two sections, we highlight in more detail the science made possible by detecting and characterizing the 21-cm background before reionization, including its fluctuations and monopole signature, and the advances needed to make this a reality in the next decade.

%%
% CONTEXT
%%
\vspace{-12pt}
\section{Scientific Context} \label{sec:context} \vspace{-12pt}
The 21-cm line from neutral hydrogen atoms has long been recognized as a powerful probe of the temperature and ionization state of the IGM before reionization \citep{Madau1997,Shaver1999}. Several radio observatories have come online in the last decade hoping to detect this background radiation, either via its spatial fluctuations \citep{Furlanetto2004,Barkana2005,Pritchard2007} or monopole spectral signature \citep[the ``global'' signal;][]{Shaver1999,Furlanetto2006}. During reionization, the prospects for jointly constraining galaxies via surveys and 21-cm observations are strong (see white paper by Furlanetto et al.), though beyond $z \sim 10$ galaxy surveys will find only the brightest galaxies \citep[e.g.,][]{Tacchella2018}, which are unlikely to be representative of the population as a whole. As a result, constraints on the first generations of stars and BHs must be sought via other probes.

The 21-cm background fundamentally probes three quantities: the density of hydrogen gas, $\delta$, the fraction of it that is neutral, $\xHI$, and its spin temperature, $\TS$, which quantifies the relative number of atoms in each hyperfine state. The observable signature is the ``differential brightness temperature,'' a measure of the HI brightness temperature relative to the background, $\TR$, generally assumed to be the CMB. Neglecting RSD, this can be written
\begin{equation}
    \dTb \simeq 27 \xHI (1 + \delta) \left(\frac{\Obnow h^2}{0.023} \right) \left(\frac{0.15}{\Omnow h^2} \frac{1 + z}{10} \right)^{1/2} \left(1 - \frac{\TR}{\TS} \right) . \label{eq:dTb}
\end{equation}
The leading factors are intuitive: a denser, more neutral medium generates stronger signals. Note, however, that if $\TS < \TR$, the signal becomes \textit{negative}, and potentially very large as $\TS \rightarrow 0$. In typical models the evolution of $\xHI$ traces the build-up of the stellar contents of galaxies, as UV photons from stars have short mean free paths and thus carve out relatively sharp HII regions, giving rise to the standard ``Swiss cheese'' picture of reionization. $\TS$, on the other hand, probes X-ray sources, which heat the gas, and sources of non-ionizing UV emission, which affect hyperfine level populations through the Wouthuysen-Field mechanism \citep{Wouthuysen1952,Field1958}. The global 21-cm signal thus provides an integral constraint on the production of UV and X-ray photons as a function of time, while spatial fluctuations in the 21-cm background constrain the masses of the typical UV and X-ray-emitting galaxies and their characteristic spectrum \citep{Pacucci2014}. This sensitivity allows 21-cm observations to address a slew of long-standing questions in galaxy formation theory, which we highlight below.

%Level populations in the hydrogen atom are set by collisions, and thus depend on the density and kinetic temperature of the gas, but can be modified by $\Lya$ photons scattering through the IGM, which ``flip'' the hyperfine levels through the Wouthuysen-Field mechanism \citep{Wouthuysen1952,Field1958}. Repeated scatterings bring $\TS$ into equilibrium with the gas kinetic temperature \citep{Chen2004,Chuzhoy2006,FurlanettoPritchard2006,Pritchard2006}, $\TK$, which is set largely by the abundance of X-ray sources, like compact objects and supernova remnants, whose emissions travels great distances before being absorbed. As a result, $\dTb$ traces the H-ionizing spectrum ($13.6 \leq h\nu /\mathrm{eV} \leq 24.6$) of sources through $\xHI$, the soft-UV spectrum ($10.2 \leq h\nu /\mathrm{eV} \leq 13.6$) through $\TS$, and the X-ray spectrum $h\nu >> 24.6 \ \mathrm{eV}$, also through $\TS$.

\begin{figure*}
\begin{center}
\includegraphics[width=0.98\textwidth]{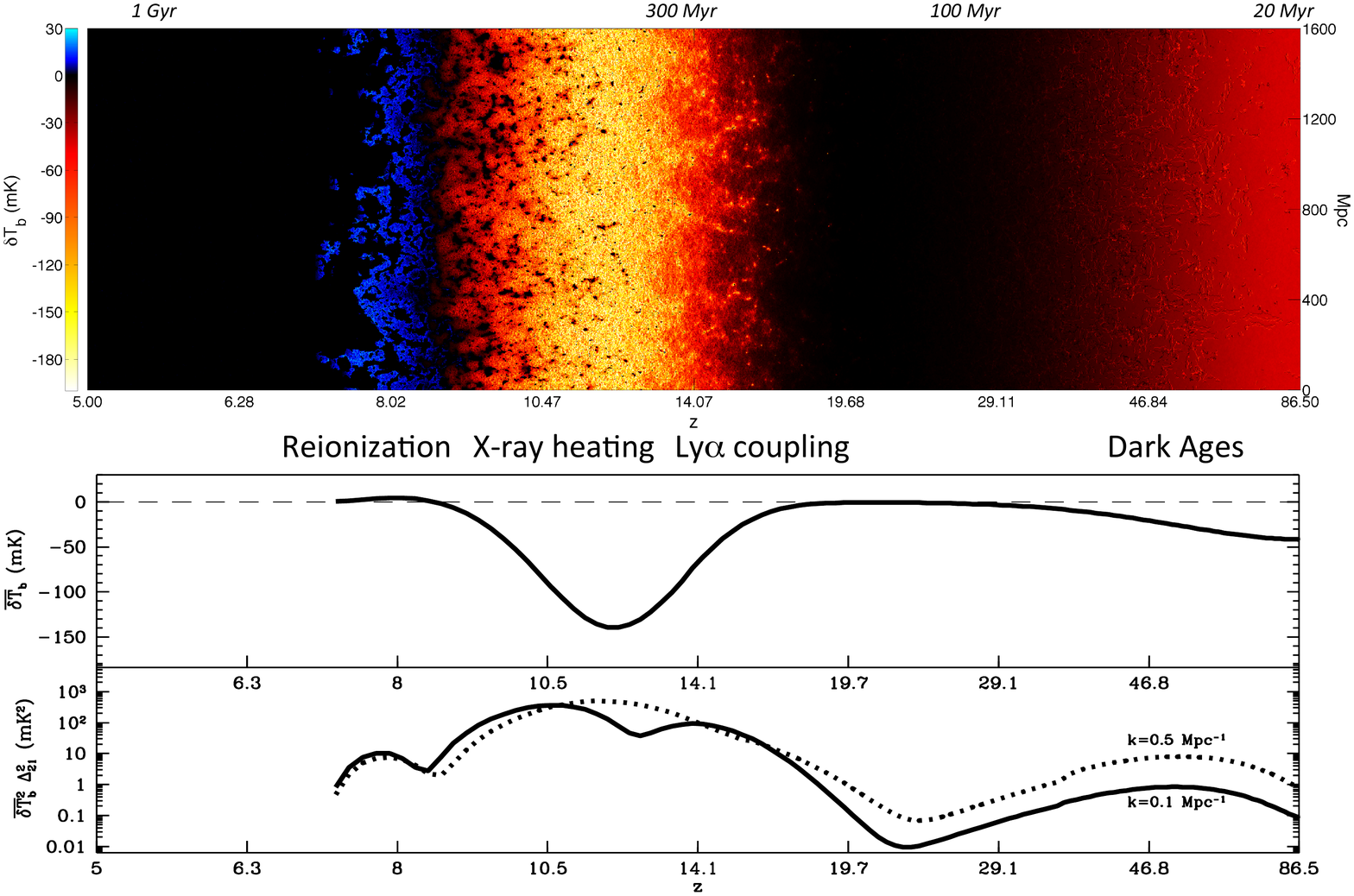}
\caption{{\bf Predictions for the evolution of the 21-cm background from \textsc{21cmfast} \cite{Mesinger2011,Mesinger2016}.} \textit{Top:} Two-dimensional slice through the simulation box showing the 21-cm brightness temperature field. The color-bar is normalized so that 21-cm absorption appears in reddish hues, emission in blue, and fully-ionized (or $\TS$ $\approx$ $\TR$) in black. \textit{Middle:} Time evolution of the global 21-cm signal, i.e., the volume-averaged 21-cm brightness temperature in the simulation box. \textit{Bottom:} Power spectrum of 21-cm fluctuations on two spatial scales as a function of redshift. }
\label{fig:gs_ps_lc}
\end{center}
\end{figure*}

%% FIRST STARS
\vspace{10pt}
\noindent {\bf When did the first stars form, and what were they like?}

%\noindent At very high redshifts, before the first luminous sources form, the evolution of the 21-cm background in a $\Lambda$CDM cosmology is very well understood. As the Universe expands the CMB cools and the density declines, giving rise to a weak absorption feature in the 21-cm background during the cosmic ``dark ages'' at $z\sim 80$ (see ``Dark Ages'' in Figure \ref{fig:gs_ps_lc}), a byproduct of successive failings of the Compton scattering and collisional coupling as mediators of the hyperfine level populations ({\color{red} refs}). At $z \lesssim 30$, the global 21-cm signal vanishes and fluctuations trace only the density field. Departures from these predictions require astrophysical sources, giving a clear signature of when stars first illuminate the cosmos.

\noindent Prior to the formation of the first stars, the global 21-cm signal is $\overline{\delta T_b} \approx 0$ and 21-cm fluctuations trace the matter field. However, once stars form, they flood the IGM with UV photons and trigger widespread Wouthuysen-Field coupling, driving $\TS$ toward the cool temperatures of the adiabatically-cooling IGM (see ``Ly-$\alpha$ coupling'' in Figure \ref{fig:gs_ps_lc}). Even knowing the redshift of this transition constrains first-star models as halos of different masses become abundant at different times. For example, the EDGES $z\sim 17$ absorption trough indicates efficient star formation in $\sim 10^9 \ M_{\odot}$ dark matter halos \citep{Mirocha2019,Kaurov2018}. The timing of this transition is also sensitive to the relative velocity between baryons and dark matter after recombination \citep{Tseliakhovich2010}, which slows accretion of gas and can thus delay the onset of star formation \citep{Tseliakhovich2011,Fialkov2012,McQuinn2012}. There is of course a degeneracy with the efficiency of UV photon production, though this could be broken with measurements of 21-cm fluctuations, which effectively isolate the characteristic mass of emitting objects through their clustering signature \citep{Furlanetto2006b}. 

%% FIRST BHs
\vspace{10pt}
\noindent {\bf Are the remnants of the first stars the seeds of today's super-massive black holes?}

\noindent  The first stars are expected to leave behind BH remnants, which form either via direct-collapse (expected for metal-free stars in the $40 \lesssim M/M_{\odot} \lesssim 120$ mass range) or as the standard end-product of type II supernovae. If at least some PopIII stars form in binaries, systems resembling high-mass X-ray binaries (HMXBs) should appear shortly after the first stars form. With relatively hard spectra, such sources are expected to begin heating the IGM on large scales, driving the mean temperature of the IGM back toward the CMB (see ``X-ray heating'' in Fig \ref{fig:gs_ps_lc}). Unique signatures of such objects may exist at redshifted 21-cm wavelengths, as the characteristic rise-and-fall of PopIII star formation \citep[see, e.g.,][]{Trenti2009,Mebane2018} will be mirrored in both the early UV and X-ray backgrounds, which imparts an asymmetry in the global 21-cm absorption signal \citep{Mirocha2018}.

One of the biggest mysteries in galaxy formation theory concerns the origins of super-massive BHs, which are known to be in place with $M \sim 10^9 \ M_{\odot}$ already at $z \gtrsim 7$ \citep{Fan2006,Banados2018,Mortlock2011}. If PopIII remnants are the progenitors of such objects they must grow by $\sim 8$ orders of magnitude in mass in less than a Gyr, implying frequent mergers or efficient accretion. The latter option would generate copius X-rays, which will affect the IGM temperature and thus the 21-cm background. Distinguishing heating by HMXBs and SMBHs may even be possible, as HMXBs form at a rate proportional to the star formation rate density (SFRD), but SMBHs -- even if formed at a rate proportional to the SFRD -- grow \textit{exponentially}, resulting in a more rapid heating rate, and thus more rapidly evolving 21-cm signal, than is expected from stellar sources \citep{EwallWice2019}. Again, detection of both the global 21-cm signal and fluctuations in the 21-cm background in principle helps break degeneracies between the characteristic host mass and accretion efficiency of high-$z$ BHs.

Understanding the formation rate of these BHs and the types of halos they form in could help establish the timing of when BH-galaxy scaling relations first emerge. In addition, constraints on the earliest BHs will provide context for the $\sim 30-40 M_{\odot}$ BH-BH mergers detected by LIGO, and thus guide predictions for high-$z$ merger rates and the stochastic gravitational wave background generated in the early Universe. 

%% LARGE SCALE FEEDBACK
\vspace{10pt}
\noindent {\bf How do large-scale UV and X-ray background exert feedback on galaxy formation?}

\noindent The emergence of large-scale radiation backgrounds is expected to impact galaxy formation in two main ways. First, Lyman-Werner (LW) photons ($11.2 \lesssim h \nu / \mathrm{eV} < 13.6$) dissociate molecular hydrogen, the primary coolant in small halos at very high redshift that have yet to be polluted with metals. The formation of the first stars -- which emit copius LW radiation -- is thus expected to restrict star formation to more massive halos capable of atomic hydrogen (or metal) line-cooling \citep{Haiman1997}. Second, Lyman-continuum (LyC) photons ($13.6 < h \nu / \mathrm{eV} < 54.4$) ionize hydrogen and helium, resulting in the growth of hot bubbles around galaxies and galaxy groups. This is expected to suppress star formation in small halos within those bubbles, whose gravitational potential wells are too shallow to retain hot gas \citep{Couchman1986}.

21-cm observations will be critical in understanding both of these feedback mechanisms. Because some photons in the LW band eventually redshift (or cascade) through the $\Lya$ line, constraints on the spin temperature at very high redshift indirectly probe the LW background. In addition, the 21-cm background constrains the growth of bubbles \textit{directly}, which opens up the possibility to study galaxies in different bubbles during reionization. However, to isolate the effects of reionization on galaxy formation from other processes, it will be vital to have constraints on the properties of galaxies \textit{before reionization begins}. 21-cm observations are likely the only way to probe the small galaxies that are most susceptible to radiative feedback, both before and during reionization, and thus has tremendous implications for the purported ``turn-over'' in the galaxy luminosity function at high-$z$ and the photon budget for reionization.

%% TRANSITION TO "BUSINESS AS USUAL"
%\vspace{10pt}
%\noindent {\bf When does ``normal'' galaxy evolution begin?}
%
%\noindent Predictions for the duration of the PopIII phase, both within individual galaxies and globally, span a wide range of possibilities. Depending on the PopIII stellar initial mass function and the efficiency with which metals mix into the interstellar medium (among other factors), PopIII stars could constitute only the initial burst of star formation in the first collapsing halos, or could continue forming throughout reionization. In some models, PopIII and PopII stellar populations co-exist in galaxies as late as $z \sim 7$. The degree to which PopIII sources are required to fit forthcoming 21-cm measurements will help to interpret limits on the duration of reionization from CMB experiments, set expectations for the detectability of PopIII sources directly at later times, and potentially provide a guide for where such sources are most likely to be found. 

%%
% MILESTONES, i.e., steps for the next decade
%%
\section{Advances Needed in the Next Decade} \label{sec:advances} \vspace{-12pt}
The high redshifts of interest for first star and BH science, $10 \lesssim z \lesssim 30$, correspond to redshifted 21-cm signals at very low radio frequencies, $45 \lesssim \nu / \mathrm{MHz} \lesssim 130$. For interferometric experiments, the scales of interest, as in the case of reionization, are $\sim 1-10$ Mpc, corresponding to angular scales of $\sim 10$-$100$ arcminutes (wavenumbers $\sim 0.1$-$10$ $h/\mathrm{Mpc}$). As a result, surveys must cover $\sim$ hundreds of square degrees, at least, in order to minimize cosmic variance. Even larger fields are advantageous for cosmological applications, while finer angular resolution would help constrain certain classes of models. For example, because fluctuations during cosmic dawn are driven by photons with long mean-free paths, we do not expect features in the 21-cm field to be as sharp as those corresponding to the edges of ionized bubbles at later times. However, if soft X-ray sources exist at $z \gtrsim 10$, e.g., intermediate or super-massive BHs whose spectra are softer than that of stellar mass BHs powering X-ray binaries, they could drive small-scale gradients in the $\delta T_b$ field. Imaging measurements with minimal instrumental effects would provide the most reliable and useful tests of these scenarios.

While imaging the 21-cm field at high resolution ($\sim$ arcminute) over large areas ($\gtrsim 100 \ \mathrm{deg}^2$) is the ultimate goal for interferometers, continued efforts targeting the global 21-cm signal are imperative as well, especially in light of the recent $\sim 78$ MHz signal reported by EDGES. Whereas interferometric campaigns can in principle avoid foreground contamination, which inhabits a well-defined region in Fourier space \citep[the ``wedge''; e.g.,][]{Datta2010,Liu2014}, for single-element receivers targeting the global signal there is no escape from the foreground -- their angular resolution is too poor to use the spatial structure in the foreground to distinguish it from the background signal. Instead, separation is performed in the frequency domain, where any spectral features introduced by the instrument could masquerade as the spectral signal of interest. Independent efforts with different designs and operating from different observing sites are essential as they suffer from different systematics and probe different sightlines through the galaxy, and thus minimize the possibility that the instrument and foreground conspire to produce an EDGES-like signal.

Regardless of whether or not an anomalous signal persists, efforts to understand the cosmic dawn are only just beginning. As experimentalists strive to detect the 21-cm power spectrum and follow-up the EDGES result, theorists must work in parallel to accommodate alternative cosmological and astrophysical models, which will help identify signatures of exotic physics and the degeneracies between astrophysics and cosmology at the high-$z$ frontier. Though there is still much work to be done in modeling the global signal and 21-cm power spectrum, on longer timescales models and inference pipelines must be adapted to work with images. Doing so will not only maximize the science return from 21-cm experiments, but enable cross-correlations with other intensity-mapping experiments and galaxy surveys, and thus guarantee progress in high-$z$ galaxy evolution and reionization for years to come.

\pagebreak

\bibliography{references}
\bibliographystyle{plain}

\end{document}

%% file: macros.tex
\newcommand{\Omnow}{\Omega_{\text{m},0}}
\newcommand{\Obnow}{\Omega_{\text{b},0}}

\newcommand{\xHI}{x_{\text{H } \textsc{i}}}

\newcommand{\Lya}{\text{Ly}\alpha}
\newcommand{\TS}{T_{\text{S}}}

\newcommand{\TR}{T_{\text{R}}}

\newcommand{\dTb}{\delta T_b}

%% file: ms.bbl
\newcommand{\noop}[1]{}
\begin{thebibliography}{10}

\bibitem{Banados2018}
E.~{Ba{\~n}ados}, B.~P. {Venemans}, C.~{Mazzucchelli}, E.~P. {Farina},
  F.~{Walter}, F.~{Wang}, R.~{Decarli}, D.~{Stern}, X.~{Fan}, F.~B. {Davies},
  J.~F. {Hennawi}, R.~A. {Simcoe}, M.~L. {Turner}, H.-W. {Rix}, J.~{Yang},
  D.~D. {Kelson}, G.~C. {Rudie}, and J.~M. {Winters}.
\newblock {An 800-million-solar-mass black hole in a significantly neutral
  Universe at a redshift of 7.5}.
\newblock {\em \nat}, 553:473--476, January 2018.

\bibitem{Barkana2018}
Rennan Barkana.
\newblock {Possible interaction between baryons and dark-matter particles
  revealed by the first stars}.
\newblock {\em Nature}, 555(7694):71--74, February 2018.

\bibitem{Barkana2005}
Rennan Barkana and Abraham Loeb.
\newblock {Detecting the earliest galaxies through two new sources of 21
  centimeter fluctuations}.
\newblock {\em \apj}, 626(1):1, 2005.

\bibitem{Berlin2018}
A.~{Berlin}, D.~{Hooper}, G.~{Krnjaic}, and S.~D. {McDermott}.
\newblock {Severely Constraining Dark-Matter Interpretations of the 21-cm
  Anomaly}.
\newblock {\em Physical Review Letters}, 121(1):011102, July 2018.

\bibitem{Bouwens2015c}
R.~J. {Bouwens}, G.~D. {Illingworth}, P.~A. {Oesch}, J.~{Caruana},
  B.~{Holwerda}, R.~{Smit}, and S.~{Wilkins}.
\newblock {Reionization After Planck: The Derived Growth of the Cosmic Ionizing
  Emissivity Now Matches the Growth of the Galaxy UV Luminosity Density}.
\newblock {\em \apj}, 811:140, October 2015.

\bibitem{Bowman2018}
Judd~D Bowman, Alan E~E Rogers, Raul~A Monsalve, Thomas~J Mozdzen, and Nivedita
  Mahesh.
\newblock {An absorption profile centred at 78 megahertz in the sky-averaged
  spectrum}.
\newblock {\em Nature}, 555(7694):67--70, March 2018.

\bibitem{Couchman1986}
H.~M.~P. {Couchman} and M.~J. {Rees}.
\newblock {Pregalactic evolution in cosmologies with cold dark matter}.
\newblock {\em \mnras}, 221:53--62, July 1986.

\bibitem{Datta2010}
A.~{Datta}, J.~D. {Bowman}, and C.~L. {Carilli}.
\newblock {Bright Source Subtraction Requirements for Redshifted 21 cm
  Measurements}.
\newblock {\em \apj}, 724:526--538, November 2010.

\bibitem{EwallWice2019}
A~Ewall-Wice, T~C Chang, and J~Lazio.
\newblock {The Radio Scream at Cosmic Dawn: A Semi-Analytic Model for the
  Impact of Radio Loud Black-Holes on the 21\,cm Global Signal}.
\newblock {\em To be submitted to AAS Journals}, March 2019.

\bibitem{EwallWice2018}
A.~{Ewall-Wice}, T.-C. {Chang}, J.~{Lazio}, O.~{Dor{\'e}}, M.~{Seiffert}, and
  R.~A. {Monsalve}.
\newblock {Modeling the Radio Background from the First Black Holes at Cosmic
  Dawn: Implications for the 21 cm Absorption Amplitude}.
\newblock {\em astro-ph/1803.01815}, March 2018.

\bibitem{Fan2006}
Xiaohui Fan.
\newblock {Evolution of high-redshift quasars}.
\newblock {\em New Astronomy Reviews}, 50(9-10):665--671, November 2006.

\bibitem{Feng2018}
C.~{Feng} and G.~{Holder}.
\newblock {Enhanced Global Signal of Neutral Hydrogen Due to Excess Radiation
  at Cosmic Dawn}.
\newblock {\em ApJ Letters}, 858:L17, May 2018.

\bibitem{Fernandez2013}
Elizabeth~R Fernandez and Saleem Zaroubi.
\newblock {The End of an Era - The Population III to Population II Transition
  and the Near Infrared Background}.
\newblock {\em arXiv.org}, (3):2047--2053, January 2013.

\bibitem{Fialkov2012}
Anastasia Fialkov, Rennan Barkana, Dmitriy Tseliakhovich, and Christopher~M
  Hirata.
\newblock {Impact of the relative motion between the dark matter and baryons on
  the first stars: semi-analytical modelling}.
\newblock {\em Monthly Notices of the Royal Astronomical Society},
  424(2):1335--1345, June 2012.

\bibitem{Field1958}
G.~B. {Field}.
\newblock {Excitation of the Hydrogen 21-CM Line}.
\newblock {\em Proceedings of the IRE}, 46:240--250, January 1958.

\bibitem{Fraser2018}
S.~{Fraser}, A.~{Hektor}, G.~{H{\"u}tsi}, K.~{Kannike}, C.~{Marzo},
  L.~{Marzola}, A.~{Racioppi}, M.~{Raidal}, C.~{Spethmann}, V.~{Vaskonen}, and
  H.~{Veerm{\"a}e}.
\newblock {The EDGES 21 cm anomaly and properties of dark matter}.
\newblock {\em Physics Letters B}, 785:159--164, October 2018.

\bibitem{Furlanetto2004}
S.~R. {Furlanetto}, M.~{Zaldarriaga}, and L.~{Hernquist}.
\newblock {The Growth of H II Regions During Reionization}.
\newblock {\em \apj}, 613:1--15, September 2004.

\bibitem{Furlanetto2006}
Steven~R Furlanetto.
\newblock {The global 21-centimeter background from high redshifts}.
\newblock {\em \mnras}, 371(2):867--878, September 2006.

\bibitem{Furlanetto2006b}
Steven~R Furlanetto, Matthew Mcquinn, and Lars Hernquist.
\newblock {Characteristic scales during reionization}.
\newblock {\em Monthly Notices of the Royal Astronomical Society},
  365(1):115--126, January 2006.

\bibitem{Haiman1997}
Z.~{Haiman}, M.~J. {Rees}, and A.~{Loeb}.
\newblock {Destruction of Molecular Hydrogen during Cosmological Reionization}.
\newblock {\em \apj}, 476:458--463, February 1997.

\bibitem{Harker2012}
Geraint J~A Harker, Jonathan~R Pritchard, Jack~O Burns, and Judd~D Bowman.
\newblock {An MCMC approach to extracting the global 21-cm signal during the
  cosmic dawn from sky-averaged radio observations}.
\newblock {\em Monthly Notices of the Royal Astronomical Society},
  419(2):1070--1084, October 2011.

\bibitem{Hill2018}
J.~C. {Hill} and E.~J. {Baxter}.
\newblock {Can early dark energy explain EDGES?}
\newblock {\em Journal of Cosmology and Astroparticle Physics}, 8:037, August
  2018.

\bibitem{Kaurov2018}
A.~A. {Kaurov}, T.~{Venumadhav}, L.~{Dai}, and M.~{Zaldarriaga}.
\newblock {Implication of the Shape of the EDGES Signal for the 21 cm Power
  Spectrum}.
\newblock {\em ApJ Letters}, 864:L15, September 2018.

\bibitem{Kovetz2018}
E.~D. {Kovetz}, V.~{Poulin}, V.~{Gluscevic}, K.~K. {Boddy}, R.~{Barkana}, and
  M.~{Kamionkowski}.
\newblock {Tighter limits on dark matter explanations of the anomalous EDGES 21
  cm signal}.
\newblock {\em Physical Review D}, 98(10):103529, November 2018.

\bibitem{Lidz2018}
A.~{Lidz} and L.~{Hui}.
\newblock {Implications of a prereionization 21-cm absorption signal for fuzzy
  dark matter}.
\newblock {\em Physical Review D}, 98(2):023011, July 2018.

\bibitem{Liu2014}
Adrian Liu, Aaron~R Parsons, and Cathryn~M Trott.
\newblock {Epoch of reionization window. I. Mathematical formalism}.
\newblock {\em Physical Review D}, 90(2):023018, July 2014.

\bibitem{Madau2018}
P.~{Madau}.
\newblock {Constraints on early star formation from the 21-cm global signal}.
\newblock {\em \mnras}, 480:L43--L47, October 2018.

\bibitem{Madau1997}
Piero Madau, Avery Meiksin, and Martin~J Rees.
\newblock {21 Centimeter Tomography of the Intergalactic Medium at High
  Redshift}.
\newblock {\em \apj}, 475:429, February 1997.

\bibitem{McQuinn2012}
M.~{McQuinn} and R.~M. {O'Leary}.
\newblock {The Impact of the Supersonic Baryon-Dark Matter Velocity Difference
  on the z \~{} 20 21 cm Background}.
\newblock {\em \apj}, 760:3, November 2012.

\bibitem{Mebane2018}
R.~H. {Mebane}, J.~{Mirocha}, and S.~R. {Furlanetto}.
\newblock {The Persistence of Population III Star Formation}.
\newblock {\em \mnras}, 479:4544--4559, October 2018.

\bibitem{Mesinger2011}
A~Mesinger, S~Furlanetto, and R~Cen.
\newblock {21cmfast: a fast, seminumerical simulation of the high-redshift
  21-cm signal - Mesinger - 2010 - Monthly Notices of the Royal Astronomical
  Society - Wiley Online Library}.
\newblock {\em \mnras}, 2011.

\bibitem{Mesinger2016}
A.~{Mesinger}, B.~{Greig}, and E.~{Sobacchi}.
\newblock {The Evolution Of 21 cm Structure (EOS): public, large-scale
  simulations of Cosmic Dawn and reionization}.
\newblock {\em \mnras}, 459:2342--2353, July 2016.

\bibitem{Miranda2017}
V.~{Miranda}, A.~{Lidz}, C.~H. {Heinrich}, and W.~{Hu}.
\newblock {CMB signatures of metal-free star formation and Planck 2015
  polarization data}.
\newblock {\em \mnras}, 467:4050--4056, June 2017.

\bibitem{Mirocha2019}
J.~{Mirocha} and S.~R. {Furlanetto}.
\newblock {What does the first highly redshifted 21-cm detection tell us about
  early galaxies?}
\newblock {\em \mnras}, 483:1980--1992, February 2019.

\bibitem{Mirocha2018}
J.~{Mirocha}, R.~H. {Mebane}, S.~R. {Furlanetto}, K.~{Singal}, and D.~{Trinh}.
\newblock {Unique signatures of Population III stars in the global 21-cm
  signal}.
\newblock {\em \mnras}, 478:5591--5606, August 2018.

\bibitem{Mortlock2011}
D.~J. {Mortlock}, S.~J. {Warren}, B.~P. {Venemans}, M.~{Patel}, P.~C. {Hewett},
  R.~G. {McMahon}, C.~{Simpson}, T.~{Theuns}, E.~A. {Gonz{\'a}les-Solares},
  A.~{Adamson}, S.~{Dye}, N.~C. {Hambly}, P.~{Hirst}, M.~J. {Irwin},
  E.~{Kuiper}, A.~{Lawrence}, and H.~J.~A. {R{\"o}ttgering}.
\newblock {A luminous quasar at a redshift of z = 7.085}.
\newblock {\em \nat}, 474:616--619, June 2011.

\bibitem{Munoz2018}
J.~B. {Mu{\~n}oz} and A.~{Loeb}.
\newblock {Insights on Dark Matter from Hydrogen during Cosmic Dawn}.
\newblock {\em astro-ph/1802.10094}, February 2018.

\bibitem{Pacucci2014}
F~Pacucci, A~Mesinger, S~Mineo, and A~Ferrara.
\newblock {The X-ray spectra of the first galaxies: 21 cm signatures}.
\newblock {\em Monthly Notices of the Royal Astronomical Society},
  443(1):678--686, July 2014.

\bibitem{Pospelov2018}
M.~{Pospelov}, J.~{Pradler}, J.~T. {Ruderman}, and A.~{Urbano}.
\newblock {Room for New Physics in the Rayleigh-Jeans Tail of the Cosmic
  Microwave Background}.
\newblock {\em Physical Review Letters}, 121(3):031103, July 2018.

\bibitem{Pritchard2007}
Jonathan~R Pritchard and Steven~R Furlanetto.
\newblock {21-cm fluctuations from inhomogeneous X-ray heating before
  reionization}.
\newblock {\em \mnras}, 376(4):1680--1694, April 2007.

\bibitem{Pritchard2010a}
Jonathan~R Pritchard and Abraham Loeb.
\newblock {Constraining the unexplored period between the dark ages and
  reionization with observations of the global 21 cm signal}.
\newblock {\em Physical Review D}, 82(2):23006, July 2010.

\bibitem{Robertson2015}
B.~E. {Robertson}, R.~S. {Ellis}, S.~R. {Furlanetto}, and J.~S. {Dunlop}.
\newblock {Cosmic Reionization and Early Star-forming Galaxies: A Joint
  Analysis of New Constraints from Planck and the Hubble Space Telescope}.
\newblock {\em The Astrophysical Journal Letters}, 802:L19, April 2015.

\bibitem{Safarzadeh2018}
M.~{Safarzadeh}, E.~{Scannapieco}, and A.~{Babul}.
\newblock {A Limit on the Warm Dark Matter Particle Mass from the Redshifted 21
  cm Absorption Line}.
\newblock {\em The Astrophysical Journal Letters}, 859:L18, June 2018.

\bibitem{Salvaterra2012}
R.~{Salvaterra}, F.~{Haardt}, M.~{Volonteri}, and A.~{Moretti}.
\newblock {Limits on the high redshift growth of massive black holes}.
\newblock {\em \aap}, 545:L6, September 2012.

\bibitem{Schauer2019}
A.~T.~P. {Schauer}, B.~{Liu}, and V.~{Bromm}.
\newblock {Constraining First Star Formation with 21cm-Cosmology}.
\newblock {\em arXiv e-prints}, January 2019.

\bibitem{Schneider2018}
A.~{Schneider}.
\newblock {Constraining noncold dark matter models with the global 21-cm
  signal}.
\newblock {\em Physical Review D}, 98(6):063021, September 2018.

\bibitem{Sharma2018}
P.~{Sharma}.
\newblock {Astrophysical radio background cannot explain the EDGES 21-cm
  signal: constraints from cooling of non-thermal electrons}.
\newblock {\em \mnras}, 481:L6--L10, November 2018.

\bibitem{Shaver1999}
P.~A. {Shaver}, R.~A. {Windhorst}, P.~{Madau}, and A.~G. {de Bruyn}.
\newblock {Can the reionization epoch be detected as a global signature in the
  cosmic background?}
\newblock {\em \aap}, 345:380--390, May 1999.

\bibitem{Tacchella2018}
S.~{Tacchella}, S.~{Bose}, C.~{Conroy}, D.~J. {Eisenstein}, and B.~D.
  {Johnson}.
\newblock {A Redshift-independent Efficiency Model: Star Formation and Stellar
  Masses in Dark Matter Halos at z > 4}.
\newblock {\em \apj}, 868:92, December 2018.

\bibitem{Trenti2009}
M.~{Trenti} and M.~{Stiavelli}.
\newblock {Formation Rates of Population III Stars and Chemical Enrichment of
  Halos during the Reionization Era}.
\newblock {\em \apj}, 694:879--892, April 2009.

\bibitem{Tseliakhovich2010}
D.~{Tseliakhovich} and C.~{Hirata}.
\newblock {Relative velocity of dark matter and baryonic fluids and the
  formation of the first structures}.
\newblock {\em Physical Review D}, 82(8):083520, October 2010.

\bibitem{Tseliakhovich2011}
Dmitriy Tseliakhovich, Dmitriy Tseliakhovich, Rennan Barkana, Christopher~M
  Hirata, and Christopher~M Hirata.
\newblock {Suppression and spatial variation of early galaxies and minihaloes}.
\newblock {\em Monthly Notices of the Royal Astronomical Society},
  418(2):906--915, December 2011.

\bibitem{Wouthuysen1952}
S.~A. {Wouthuysen}.
\newblock {On the excitation mechanism of the 21-cm (radio-frequency)
  interstellar hydrogen emission line.}
\newblock {\em The Astronomical Journal}, 57:31--32, 1952.

\end{thebibliography}
